\documentclass{lmcs} 
\usepackage[utf8]{inputenc}
\pdfoutput=1

\usepackage{lastpage}
\lmcsdoi{18}{2}{3}
\lmcsheading{}{\pageref{LastPage}}{}{}%
{Sep.~21,~2021}{Apr.~18,~2022}{}


\usepackage{graphicx}
\usepackage{enumitem}
\usepackage{placeins}
\usepackage{fancyvrb}
\usepackage{xcolor}
\usepackage{hhline}
\usepackage{amsmath}
\usepackage{booktabs}
\usepackage{longtable}
\usepackage{listings}
\usepackage{amssymb}
\usepackage{algorithm,algpseudocode}
\usepackage{marvosym}
\usepackage{tikz}
\usepackage{pgfplots}
\usepackage{afterpage}
\usepackage{float}

\pgfplotsset{compat=1.14}
\definecolor{drat}{HTML}{0C06F7}
\definecolor{frat}{HTML}{019529}
\definecolor{dratlight}{HTML}{A2CFFE}
\definecolor{fratlight}{HTML}{9BE5AA}

\setbox1\hbox{\verb|l|}
\newlength{\vcharwidth}
\algrenewcommand{\algorithmiccomment}[1]{\hskip3em$\rhd$ #1}
\algdef{SE}[SUBALG]{Indent}{EndIndent}{}{\algorithmicend\ }%
\algdef{SE}[SWITCH]{Case}{EndCase}[1]{\textbf{case}\ #1\ \textbf{of}}{\algorithmicend\ \algorithmicswitch}%
\algtext*{Indent}
\algtext*{EndIndent}
\algtext*{EndCase}
\algtext*{EndFor}
\algtext*{EndWhile}
\algtext*{EndIf}
\algtext*{EndFunction}

\newcommand{\FRAT}{\textsf{FRAT}}
\newcommand{\FRATZ}{\textsf{FRAT0}}
\newcommand{\DRAT}{\textsf{DRAT}}
\newcommand{\LRAT}{\textsf{LRAT}}
\newcommand{\DPR}{\textsf{DPR}}
\newcommand{\DRATtrim}{\textsf{DRAT-trim}}
\newcommand{\DPRtrim}{\textsf{DPR-trim}}
\newcommand{\PRTODRAT}{\textsf{PR2DRAT}}
\newcommand{\LPR}{\textsf{LPR}}
\newcommand{\cadical}{\textsc{Ca\-Di\-CaL}}
\newcommand{\minisat}{\textsc{Mini\-Sat}}
\definecolor{red}{rgb}{.7,.2,.1}
\definecolor{grn}{rgb}{.1,.5,.1}
\definecolor{blu}{rgb}{0,.2,.5}
\definecolor{yel}{rgb}{.5,.5,.15}

\begin{document}
\title{A Flexible Proof Format for SAT Solver-Elaborator Communication}

\author[S.~Baek]{Seulkee Baek\rsuper{*}}	
\thanks{\textsuperscript{*\phantom{*}} Partially supported by AFOSR grant FA9550--18--1--0120}	

\author[M.~Carneiro]{Mario Carneiro}	

\author[M.J.H.~Heule]{Marijn J.H. Heule\rsuper{**}}	
\thanks{\textsuperscript{**} Supported by the National Science Foundation under grant CCF-2010951}	

\address{Carnegie Mellon University, Pittsburgh, PA, United States}	
\email{\{seulkeeb,mcarneir,mheule\}@andrew.cmu.edu}  

\begin{abstract}
We introduce \FRAT, a new proof format for unsatisfiable SAT problems, and its associated toolchain. Compared to \DRAT, the \FRAT\ format allows solvers to include more information in proofs to reduce the computational cost of subsequent elaboration to \LRAT\@. The format is easy to parse forward and backward, and it is extensible to future proof methods. The provision of optional proof steps allows SAT solver developers to balance implementation effort against elaboration time, with little to no overhead on solver time. We benchmark our \FRAT\ toolchain against a comparable \DRAT\ toolchain and confirm $>$84\% median reduction in elaboration time and $>$94\% median decrease in peak memory usage.
\end{abstract}

\maketitle              

\section{Introduction}

The \textit{Boolean satisfiability problem} is the problem of determining, for a given Boolean formula consisting of Boolean variables and connectives, whether there exists a variable assignment under which the formula evaluates to true. Boolean satisfiability (SAT) is interesting in part because there are surprisingly diverse types of problems that can be encoded as Boolean formulas and solved efficiently by checking their satisfiability. \textit{SAT solvers}, programs that automatically solve SAT problems, have been successfully applied to a wide range of areas, including hardware verification~\cite{biere1999symbolic}, planning~\cite{kautz1996pushing}, and combinatorics~\cite{heule2016solving}.

The performance of SAT solvers has taken great strides in recent years, and modern solvers can often solve problems involving millions of variables and clauses, which would have been unthinkable a mere 20 years ago~\cite{knuth2015art}. But this improvement comes at the cost of significant increase in the code complexity of SAT solvers, which makes it difficult to either assume their correctness on faith, or certify their program correctness directly. As a result, the ability of SAT solvers to produce independently verifiable certificates has become a pressing necessity. Since there is an obvious certificate format (the satisfying Boolean assignment) for satisfiable problems, the real challenge in proof-producing SAT solving is in devising a compact proof format for unsatisfiable problems, and developing a toolchain that efficiently produces and verifies it.

The current de facto standard proof format for unsatisfiable SAT problems is \DRAT~\cite{drat}. The format, as well as its predecessor DRUP, were
designed with a strong focus on quick adaptation by the community, emphasizing easy proof emission, practically zero overhead, and reasonable validation speed~\cite{bridge}. The \DRAT\ format has become the only supported proof format in SAT Competitions and Races since 2014 due to entrants losing interest in alternatives.

\DRAT\ is a \textit{clausal} proof format~\cite{goldberg2003verification}, which means that a \DRAT\ proof consists of a sequence of instructions for adding and deleting clauses. It is helpful to think of a \DRAT\ proof as a program for modifying the `active multiset' of clauses: the initial active multiset is the clauses of the input problem, and this multiset grows and shrinks over time as the program is executed step by step. The invariant throughout program execution is that the active multiset at any point of time is \textit{at least as satisfiable} as the initial active multiset. This invariant holds trivially in the beginning and after a deletion; it is also preserved by addition steps by either RUP or RAT, which we explain shortly. The last step of a \DRAT\ proof is the addition of the empty clause, which ensures the unsatisfiability of the final active multiset, and hence that of the initial active multiset, i.e.\ the input problem.

Every addition step in \DRAT\ is either a \emph{reverse unit propagation} (RUP) step~\cite{goldberg2003verification} or a \emph{resolution asymmetric tautology} (RAT)~\cite{rules} step. A clause $C$ has the property \textsf{AT} (asymmetric tautology) with respect to a formula $F$ if $F,\overline{C}\vdash_1\bot$, which is to say, there is a proof of the empty clause by unit propagation using $F$ and the negated literals in $C$. A RUP step that adds $C$ to the active multiset $F$ is valid if $C$ has property \textsf{AT} with respect to $F$. A clause $l\lor C$ has property \textsf{RAT} with respect to $F$ if for every clause $\overline{l}\lor D\in F$, the clause $C\lor D$ has property \textsf{AT} with respect to $F$. In this case, $C$ is not logically entailed by $F$, but $F$ and $F\wedge C$ are equisatisfiable, and a RAT step will add $C$ to the active multiset if $C$ has property \textsf{RAT} with respect to $F$. (See~\cite{drat} for more about the justification for this proof system.)

\DRAT\ has a number of advantages over formats based on more traditional proof calculi, such as resolution or analytic tableaux. For SAT solvers, \DRAT\ proofs are easier to emit because CNF clauses are the native data structures that the solvers store and manipulate internally. Whenever a solver obtains a new clause, the clause can be simply streamed out to a proof file without any further modification. Also, \DRAT\ proofs are more compact than resolution proofs, as the latter can become infeasibly large for some classes of SAT problems~\cite{pigeonhole}.

There is, however, room for further improvement in the \DRAT\ format due to the information loss incurred by \DRAT\ proofs. Consider, for instance, the SAT problem and proofs shown in Figure~\ref{fig:drat-lrat}. The left column is the input problem in the DIMACS format, the center column is its \DRAT\ proof, and the right column is the equivalent proof in the \LRAT~\cite{lrat} format, which can be thought of as an enriched version of \DRAT\ with more information. The numbers before the first zero on lines without a ``\texttt{d}'' represent literals: positive numbers denote positive literals, while negative numbers denote negative literals. The first clause of the input formula is $(x_1 \lor x_2 \lor \overline x_3)$, or equivalently {\tt \textcolor{grn}{1 2 -3} 0} in DIMACS\@.

The first lines of both \DRAT\ and \LRAT\ proofs are RUP steps for adding the clause $(x_1 \lor x_2)$, written \texttt{\textcolor{red}{1 2} 0}. When an \LRAT\ checker verifies this step, it is informed of the IDs of active clauses (the trailing numbers \texttt{\textcolor{yel}{1 6 3}}) relevant for unit propagation, in the exact order they should be used. Therefore, the \LRAT\ checker only has to visit the first, sixth, and third clauses and confirm that, starting with unit literals $\overline{x_1}, \overline{x_2}$, they yield the new unit literals $\overline{x_3}, x_4, \bot$. In contrast, a \DRAT\ checker verifying the same step must add the literals $\overline{x_1}, \overline{x_2}$ to the active multiset (in this case, the eight initial clauses) and carry out a blind unit propagation with the whole resulting multiset until reaching contradiction. This omission of RUP information in \DRAT\ proofs introduces significant overheads in proof verification. Although the exact figures vary from problem to problem, checking a \DRAT\ proof typically takes approximately twice as long as solving the original problem, whereas the verification time for an \LRAT\ proof is negligible compared to its solution time. This additional cost of checking \DRAT\ proofs also represents a lost opportunity: when a SAT solver emits a RUP step, it knows exactly how the new clause was obtained, and this knowledge can (in theory) be turned into an \LRAT-style RUP annotation, which can cut down verification costs significantly if conveyed to the verifier.

For the \DRAT\ format, a design choice was made not to include such information since demanding explicit proofs for all steps turned out to be impractical. Although it is \textit{theoretically} possible to always glean the correct RUP annotation from the solver state, computing this information can be intricate and costly for some types of inferences (e.g.\ conflict-clause minimization~\cite{minimization}), making it harder to support proof logging~\cite{AVG09}. Reducing such overheads is particularly important for solving satisfiable formulas, as proofs are superfluous for them and the penalty for maintaining such proofs should be minimized. We should note, however, that proof elaboration need not be an all-or-nothing business; if it is infeasible to demand 100\% elaborated proofs, we can still ask solvers to fill in as many gaps as it is convenient for them to do so, which would still be a considerable improvement over handling all of it from the verifier side.

\begin{figure}[tb]
\centering
\begin{tabular}{c@{\hskip 0.5in}c@{\hskip 0.5in}c}
 \textsf{DIMACS} & \textsf{DRAT} & \textsf{LRAT} \\ [2ex]
\begin{minipage}{10\vcharwidth}
\begin{Verbatim}[commandchars=\\\{\}]
p cnf 4 8
\textcolor{grn}{ 1  2 -3} 0
\textcolor{grn}{-1 -2  3} 0
\textcolor{grn}{ 2  3 -4} 0
\textcolor{grn}{-2 -3  4} 0
\textcolor{grn}{-1 -3 -4} 0
\textcolor{grn}{ 1  3  4} 0
\textcolor{grn}{-1  2  4} 0
\textcolor{grn}{ 1 -2 -4} 0



\end{Verbatim}
\end{minipage}
&
\begin{minipage}{11\vcharwidth}
\begin{Verbatim}[commandchars=\\\{\}]
 \textcolor{red}{     1 2} 0
d\textcolor{blu}{  1 -3 2} 0
 \textcolor{red}{     1 3} 0
d\textcolor{blu}{   1 4 3} 0
 \textcolor{red}{       1} 0
d\textcolor{blu}{     1 3} 0
d\textcolor{blu}{     1 2} 0
d\textcolor{blu}{ 1 -4 -2} 0
 \textcolor{red}{       2} 0
d\textcolor{blu}{  -1 4 2} 0
d\textcolor{blu}{  2 -4 3} 0
 \textcolor{red}{        } 0
\end{Verbatim}
\end{minipage}
&
\begin{minipage}{22\vcharwidth}
\begin{Verbatim}[commandchars=\\\{\}]
\textcolor{blu}{ 9}\textcolor{red}{ 1 2} 0 \textcolor{yel}{      1 6 3} 0
\textcolor{blu}{ 9}               d \textcolor{blu}{1} 0
\textcolor{blu}{10}\textcolor{red}{ 1 3} 0 \textcolor{yel}{      9 8 6} 0
\textcolor{blu}{10}               d \textcolor{blu}{6} 0
\textcolor{blu}{11}\textcolor{red}{   1} 0 \textcolor{yel}{   10 9 4 8} 0
\textcolor{blu}{11}          d \textcolor{blu}{10}
                 \textcolor{blu}{9}
                   \textcolor{blu}{8} 0
\textcolor{blu}{12}\textcolor{red}{   2} 0 \textcolor{yel}{   11 7 5 3} 0
\textcolor{blu}{12}             d \textcolor{blu}{7}
                   \textcolor{blu}{3} 0
\textcolor{blu}{13}\textcolor{red}{    } 0 \textcolor{yel}{11 12 2 4 5} 0
\end{Verbatim}
\end{minipage}

\end{tabular}
\caption{\DRAT\ and \LRAT\ proofs of a SAT problem. All whitespace and alignment is not significant; we have aligned lines of the \DRAT\ proof with the corresponding \LRAT\ lines (\texttt{d} steps in \LRAT\ may correspond to multiple \DRAT\ \texttt{d} steps).}%
\label{fig:drat-lrat}
\end{figure}

Inclusion of final clauses is another potential area for improvement over the \DRAT\ format. A \DRAT\ proof typically includes many addition steps that do not ultimately contribute to the derivation of the empty clause. This is unavoidable in the proof emission phase, since a SAT solver cannot know in advance whether a given clause will be ultimately useful, and must stream out the clause before it can find out. All such steps, however, should be dropped in the post-processing phase in order to compress proofs and speed up verification. The most straightforward way of doing this is processing the proof in reverse order~\cite{goldberg2003verification}: when processing a clause $C_{k + 1}$, identify all the clauses used to derive $C_{k + 1}$, mark them as `used', and move on to clause $C_k$. For each clause, process it if it is marked as used, and skip it otherwise. The only caveat of this method is that the postprocessor needs to know which clauses were present at the very end of the proof, since there is no way to identify which clauses were used to derive the empty clause otherwise. Although it is possible to enumerate the final clauses by a preliminary forward pass through a \DRAT\ proof, this is clearly unnecessary work since SAT solvers know exactly which clauses are present at the end, and it is desirable to put this information in the proof in the first place.

In this paper, we introduce \FRAT, a new proof format designed to address the above issues by allowing more fine-grained communication between SAT solvers and elaborators. Using a diverse range of benchmarks with nontrivial CNF instances, we also demonstrate that the \FRAT\ format achieves its design goals with significant performance improvements.

A quick note on versions: the previous version of this paper~\cite{baek2021flexible} was presented at TACAS 2021. The main changes from the previous version are the extension of \FRAT\ to accommodate PR proof steps (Section~\ref{sec:pr}) and new benchmarks for unannotated \FRAT\ proofs (Section~\ref{sec:fratz-vs-drat}) and \DPR\ proofs (Sections~\ref{sec:frat-elab-vs-dpr-trim} and~\ref{sec:frat-elab-vs-pr2drat}).

\section{The \FRAT\ format}\label{sec:frat}

The design of the \FRAT\ format is largely based on \DRAT, with which it shares many similarities. The main differences between \FRAT\ and \DRAT\ are:
\begin{enumerate}[label={(\arabic*)}]
\item optional annotation of RUP steps,
\item inclusion of final clauses, and
\item identification of clauses by unique IDs.
\end{enumerate}
We’ve already explained the rationale for (1) and (2); (3) is necessary for concise references to clauses in deletions and RUP step annotations. More specifically, a \FRAT\ proof consists of the following seven types of proof steps:

\begin{itemize}

\item[\texttt{o}:] An original step; a clause from the input file. The purpose of these lines is to name the clauses from the input with identifiers; they are not required to come in the same order as the file, they are not required to be numbered in order, and not all steps in the input need appear here. Proof may also progress (with \texttt{a} and \texttt{d} steps) before all \texttt{o} steps are added.

\item[\texttt{a},]\texttt{l}: An addition step, and an optional \LRAT-style unit propagation proof of the step. The proof, if provided, is a sequence of clauses in the current formula in the order that they become unit. For solver flexibility, they are allowed to come out of order, but the elaborator is optimized for the case where they are correctly ordered. For a RAT or PR step, the negative numbers in the proof refer to the clauses in the active set that contain the negated pivot literal, followed by the unit propagation proof of the resolvent. See~\cite{lrat} for more details on the \LRAT\ checking algorithm.

For PR steps (see Section~\ref{sec:pr}), the list of literals encodes both the clause $C$ and the witness $\omega$. These are subject to the constraint $C[0]=\omega[0]$, so a step has a witness iff $C[0]$ appears again later in the clause.
\item[\texttt{d}:] A deletion step for deleting the clause with the given ID from the formula. The literals given must match the literals in the corresponding addition step up to permutation.
\item[\texttt{r}:] A relocation step. The syntax is \texttt{r $\langle ids\rangle$ 0}, where $\langle ids\rangle$ has the form $s_0$, $t_0$, \dots, $s_k$, $t_k$ and must consist of an even number of clause IDs. It indicates that the active clause with ID $s_i$ is re-labeled and now has ID $t_i$, for each $0 \le i \le k$. (This is used for solvers that use pointer identity for clauses, but also move clauses during garbage collection.)
\item[\texttt{f}:] A finalization step. These steps must come at the end of a proof (after \texttt{a}, \texttt{d}, \texttt{o} steps), and provide the list of all active clauses at the end of the proof. The clauses may come in any order, but every step that has been added and not deleted must be present, and for it to be a proof of unsatisfiability an \texttt{f} step for the empty clause must exist. (For best results, clauses should be finalized in roughly reverse order of when they were added.)

\item[\texttt{c}:] A text comment, not a real step. These are preserved by the toolchain so they can be used to mark or classify proof steps for debugging purposes, but they are removed in \textsf{LRAT} output because external verifiers generally don't have support for comment lines.
\end{itemize}

\noindent
Note that our modified version of \cadical\ also outputs steps of the form $\mathtt{t}\ \langle todo\_id\rangle\ \mathtt{0}$, an outdated comment syntax now replaced by the more expressive \texttt{c} steps. The \texttt{t} steps are used to collect statistics on code paths that produce \texttt{a} steps without proofs. See Section~\ref{sec:solvers} for how this information is used.

\begin{figure}[tb]
\centering

\FRAT\\[7pt]

\begin{tabular}{c@{\hskip 1.5em}|@{\hskip 1.5em}c} 
\begin{minipage}{26\vcharwidth}
\begin{Verbatim}[commandchars=\\\{\}]
o \textcolor{blu}{1 }\textcolor{grn}{              1 2 -3} 0
o \textcolor{blu}{2 }\textcolor{grn}{             -1 -2 3} 0
o \textcolor{blu}{3 }\textcolor{grn}{              2 3 -4} 0
o \textcolor{blu}{4 }\textcolor{grn}{             -2 -3 4} 0
o \textcolor{blu}{5 }\textcolor{grn}{            -1 -3 -4} 0
o \textcolor{blu}{6 }\textcolor{grn}{               1 3 4} 0
o \textcolor{blu}{7 }\textcolor{grn}{              -1 2 4} 0
o \textcolor{blu}{8 }\textcolor{grn}{             1 -2 -4} 0
a \textcolor{blu}{9 }\textcolor{red}{-3 -4} 0 l \textcolor{yel}{     5 1 8} 0
a \textcolor{blu}{10}\textcolor{red}{   -4} 0 l \textcolor{yel}{   9 3 2 8} 0
a \textcolor{blu}{11}\textcolor{red}{    3} 0
a \textcolor{blu}{12}\textcolor{red}{   -2} 0
a \textcolor{blu}{13}\textcolor{red}{    1} 0 l \textcolor{yel}{   12 11 1} 0
a \textcolor{blu}{14}\textcolor{red}{     } 0 l \textcolor{yel}{13 12 10 7} 0
\end{Verbatim}
\end{minipage}
&
\begin{minipage}{14\vcharwidth}
\begin{Verbatim}[commandchars=\\\{\}]
f \textcolor{blu}{1 }\textcolor{red}{  1 2 -3} 0
f \textcolor{blu}{2 }\textcolor{red}{ -2 -1 3} 0
f \textcolor{blu}{3 }\textcolor{red}{  2 3 -4} 0
f \textcolor{blu}{4 }\textcolor{red}{ -2 -3 4} 0
f \textcolor{blu}{5 }\textcolor{red}{-1 -3 -4} 0
f \textcolor{blu}{6 }\textcolor{red}{   1 3 4} 0
f \textcolor{blu}{7 }\textcolor{red}{  -1 2 4} 0
f \textcolor{blu}{8 }\textcolor{red}{ 1 -2 -4} 0
f \textcolor{blu}{9 }\textcolor{red}{   -3 -4} 0
f \textcolor{blu}{10}\textcolor{red}{      -4} 0
f \textcolor{blu}{11}\textcolor{red}{       3} 0
f \textcolor{blu}{12}\textcolor{red}{      -2} 0
f \textcolor{blu}{13}\textcolor{red}{       1} 0
f \textcolor{blu}{14}\textcolor{red}{        } 0
\end{Verbatim}
\end{minipage}
\end{tabular}

\vspace{10pt}

\LRAT\\[7pt]

\begin{minipage}{22\vcharwidth}
\begin{Verbatim}[commandchars=\\\{\}]
\textcolor{blu}{ 9} \textcolor{red}{-3 -4} 0 \textcolor{yel}{     5 1 8} 0
\textcolor{blu}{ 9}                d \textcolor{blu}{5} 0
\textcolor{blu}{10} \textcolor{red}{   -4} 0 \textcolor{yel}{   9 3 2 8} 0
\textcolor{blu}{10}            d \textcolor{blu}{8 3 9} 0
\textcolor{blu}{11} \textcolor{red}{    3} 0 \textcolor{yel}{  10 6 7 2} 0
\textcolor{blu}{11}              d \textcolor{blu}{2 6} 0
\textcolor{blu}{12} \textcolor{red}{   -2} 0 \textcolor{yel}{   11 10 4} 0
\textcolor{blu}{12}                d \textcolor{blu}{4} 0
\textcolor{blu}{13} \textcolor{red}{    1} 0 \textcolor{yel}{   12 11 1} 0
\textcolor{blu}{13}             d \textcolor{blu}{1 11} 0
\textcolor{blu}{14} \textcolor{red}{     } 0 \textcolor{yel}{13 12 10 7} 0

\end{Verbatim}
\end{minipage}

\caption{\FRAT\ and \LRAT\ proofs of a SAT problem. To illustrate that proofs are optional, we have omitted the proofs of steps \textcolor{blu}{\texttt{11}} and \textcolor{blu}{\texttt{12}} in this example. The steps must still be legal RAT steps but the elaborator will derive the proof rather than the solver.}%
\label{fig:frat-lrat}
\end{figure}

Figure~\ref{fig:drat-lrat} is an example from~\cite{lrat}, which includes a SAT problem in DIMACS format, and the proofs of its unsatisfiability in \DRAT\ and \LRAT\ formats. It shows how proofs are produced and elaborated via the \DRAT\ toolchain. Figure~\ref{fig:frat-lrat} shows the corresponding problem and proofs for the \FRAT\ toolchain. Notice how the \FRAT\ proof is more verbose than its \DRAT\ counterpart and includes all the hints for addition steps, which are reused in the subsequent \LRAT\ proof.

\begin{figure}
    \centering
    \begin{align*}
     \langle proof \rangle &\leftarrow \langle line \rangle^* \\
     \langle line \rangle &\leftarrow \langle comment \rangle \mid \langle orig \rangle \mid \langle add \rangle \mid \langle del \rangle \mid \langle final\rangle \mid \langle reloc \rangle \\
     \langle add \rangle &\leftarrow \langle add\_seg \rangle \mid \langle add\_seg \rangle\ \langle hint \rangle\\
     \langle comment \rangle &\leftarrow \mathtt{c}\ [\texttt{\textasciicircum\textbackslash n}]^*\ \mathtt{.}\\
     \langle orig \rangle &\leftarrow \mathtt{o}\ \langle id \rangle\ \langle literal \rangle^*\ \mathtt{0}\\
     \langle add\_seg \rangle &\leftarrow \mathtt{a}\ \langle id \rangle\ \langle literal \rangle^*\ \mathtt{0} \mid {\color{red}\mathtt{a}\ \langle id \rangle\ \langle literal \rangle^*\ \langle literal \rangle^*\ \mathtt{0}}\\
     \langle del \rangle &\leftarrow \mathtt{d}\ \langle id \rangle\ \langle literal \rangle^*\ \mathtt{0}\\
     \langle final \rangle &\leftarrow \mathtt{f}\ \langle id \rangle\ \langle literal \rangle^*\ \mathtt{0}\\
     \langle reloc \rangle &\leftarrow \mathtt{r}\ (\langle id \rangle\ \langle id \rangle)^*\ \mathtt{0}\\
     \langle hint \rangle &\leftarrow \mathtt{l}\ (\langle id \rangle \mid -\langle id \rangle)^*\ \mathtt{0}\\
     \langle id \rangle &\leftarrow \langle pos \rangle\\
     \langle literal \rangle &\leftarrow \langle pos \rangle \mid \langle neg \rangle\\
     \langle neg \rangle &\leftarrow \mathtt{-}\langle pos \rangle\\
     \langle pos \rangle &\leftarrow [\texttt{1-9}]\ [\texttt{0-9}]^*
    \end{align*}
    \caption{Context-free grammar for the \FRAT\ format. (The additions in {\color{red}red} are to support PR proofs, see Section~\ref{sec:pr}.)}%
    \label{fig:frat_cfg}
\end{figure}

\subsubsection{Binary \FRAT}

The files shown in Figure~\ref{fig:frat-lrat} are in the text version of the \FRAT\ format, but for efficiency reasons solvers may also wish to use a binary encoding. The binary \FRAT\ format is exactly the same in structure, but the integers are encoded using the same variable-length integer encoding used in binary \DRAT~\cite{compression}. Unsigned numbers are encoded in 7-bit little endian, with the high bit set on each byte except the last. That is, the number \[n=x_0+2^7x_1+\cdots+2^{7k}x_k\] (with each $x_i<2^7$) is encoded as \[\mathtt{1}x_0\;\mathtt{1}x_1\;\dots\;\mathtt{0}x_k.\] Signed numbers are encoded by mapping $n\ge 0$ to $f(n):=2n$ and $-n$ (with $n>0$) to $f(n):=2n+1$, and then using the unsigned encoding. (Incidentally, the mapping $f$ is not surjective, as it misses $1$. But it is used by other formats so we have decided not to change it.)

Comments using the \texttt{c} step syntax are preserved even in the binary format; they are encoded as the character \texttt{c} followed by a null-terminated string with the contents of the comment.
\begin{figure}[htb]
\centering
\begin{tabular}{r|cccccccccc} 
text&\texttt{a}&\textcolor{blu}{\texttt{9}}&\textcolor{red}{\texttt{-3}}&\textcolor{red}{\texttt{-4}}&\texttt{0}&\texttt{l}&\textcolor{yel}{\texttt{5}}&\textcolor{yel}{\texttt{1}}&\textcolor{yel}{\texttt{8}}&\texttt{0}\\
binary&\texttt{61}&\textcolor{blu}{\texttt{09}}&\textcolor{red}{\texttt{07}}&\textcolor{red}{\texttt{09}}&\texttt{00}&\texttt{6C}&\textcolor{yel}{\texttt{0A}}&\textcolor{yel}{\texttt{02}}&\textcolor{yel}{\texttt{10}}&\texttt{00}
\end{tabular}
\caption{Comparison of binary and text formats for a step. Note that the step ID \textcolor{blu}{\texttt{9}} uses the unsigned encoding, but literals and \LRAT\ style proof steps use signed encoding.}%
\label{fig:binary_frat}
\end{figure}
\subsection{Flexibility and extensibility}

The purpose of the \FRAT\ format is for solvers to be able to quickly write down what they are doing while they are doing it, with the elaborator stage ``picking up the pieces'' and preparing the proof for consumption by simpler mechanisms such as certified \LRAT\ checkers. As such, it is important that we are able to concisely represent all manner of proof methods used by modern SAT solvers.

The high level syntax of a \FRAT\ file is quite simple: A sequence of ``segments'', each of which begins with a character, followed by zero or more nonzero numbers, followed by a \texttt{0}. In the binary version, each segment similarly begins with a printable character, followed by zero or more nonzero bytes, followed by a zero byte. (Note that continuation bytes in an unsigned number encoding are always nonzero.) This means that it is possible to jump into a \FRAT\ file and find segment boundaries by searching for a nearby zero byte.

This is in contrast to binary \LRAT, in which add steps are encoded as\\ $\mathtt{a}\ \langle id \rangle\ \langle literal \rangle^*\ \mathtt{0}\ (\pm\langle id \rangle)^*\ \mathtt{0}$, because a random zero byte could either be the end of a segment or the middle of an add step. Since \texttt{0x61}, the ASCII representation of \texttt{a}, is also a valid step ID (encoding the signed number $-48$), in a sequence such as $(\texttt{a}\ \langle nonzero\rangle^*\ \mathtt{0})^*$, the literals and the steps cannot be locally disambiguated.

The local disambiguation property is important for our \FRAT\ elaborator, because it means that we can efficiently parse \FRAT\ files generated by solvers \emph{backward}, reading the segments in reverse order so that we can perform backward checking in a single pass. The terminating `\texttt{.}' in \texttt{c} comments is a concession to ensure the backward parsing property; without this, sequences like $\texttt{c}\ (\texttt{a}\ \langle nonzero\rangle^*\ \mathtt{0})^*$ would require excessive lookahead (reading from the right) to determine that in fact all the add steps are commented out.

\DRAT\ is based on adding clauses that are \textsf{RAT} with respect to the active formula. It is quite versatile and sufficient for most common cases, covering CDCL steps, hyper-resolution, unit propagation, blocked clause elimination and many other techniques. However, we recognize that some methods can be too difficult or expensive to translate into this proof system. In this work we define only seven segment characters (\texttt{a}, \texttt{c}, \texttt{d}, \texttt{f}, \texttt{l}, \texttt{o}, \texttt{r}), that suffice to cover methods used by SAT solvers targeting \DRAT\@. However, the format is forward-compatible with new kinds of proof steps, that can be indicated with different characters.

For example, \textsc{CryptoMiniSat}~\cite{Soos09} is a SAT solver that also supports XOR clause extraction and reasoning, and can derive new XOR clauses using proof techniques such as Gaussian elimination. Encoding this in \DRAT\ is quite complicated: The XOR clauses must be Tseitin transformed into CNF, and Gaussian elimination requires a long resolution proof. Participants in SAT competitions therefore turn this reasoning method off as producing the \DRAT\ proofs is either too difficult or the performance gains are canceled out by the overhead.

\FRAT\ resolves this impasse by allowing the solver to express itself with minimal encoding overhead. A hypothetical extension to \FRAT\ would add new segment characters to allow adding and deleting XOR clauses, and a new proof method for proof by linear algebra on these clauses. The \FRAT\ elaborator would be extended to support the new step kinds, and it could either perform the expensive translation into \DRAT\ at that stage (only doing the work when it is known to be needed for the final proof), or it could pass the new methods on to some \textsf{XLRAT} backend format that understands these steps natively. Since the extension is backward compatible, it can be done without impacting any other \FRAT-producing solvers.

\subsection{Extensibility case study: PR proof steps}\label{sec:pr}

As a demonstration of its extensibility, we extended the \FRAT\ format to support the propagational redundancy (PR) steps~\cite{heule2017short} used by the \DPR\ format and \DPRtrim\ tool.

For consistency with \DPR\ proofs supported by \DPRtrim, we encode PR steps using the \texttt{a} command, as shown in Figure~\ref{fig:frat_cfg}. As written this CFG is ambiguous, since we cannot tell the difference between regular steps of the form $\mathtt{a}\ \langle id \rangle\ \langle literal \rangle^*\ \mathtt{0}$, and PR steps of the form $\mathtt{a}\ \langle id \rangle\ \langle literal \rangle^*\ \langle literal \rangle^*\ \mathtt{0}$. However, regular \texttt{a} steps have the constraint that they must not repeat any literals, while PR steps $\mathtt{a}(i,C,\omega)$ satisfy $C[0]=\omega[0]$, so to distinguish them we search for an occurrence of $C[0]$ later in the clause.

As we shall see in Section~\ref{sec:frat-elab-vs-dpr-trim}, this extension works very well in practice with no adverse effects on performance, providing further evidence of the extensibility of the \FRAT\ format to new types of inferences.

\section{\FRAT-producing solvers}\label{sec:solvers}

The \FRAT\ proof format is designed to allow conversion of \DRAT-producing solvers into \FRAT-producing solvers at minimal cost, both in terms of implementation effort and impact on runtime efficiency. In order to show the feasibility of such conversions, we chose two popular SAT solvers, \cadical\footnote{\url{https://github.com/digama0/cadical}} and \minisat\footnote{\url{https://github.com/digama0/minisat}}, to modify as case studies. The solvers were chosen to demonstrate two different aspects of feasibility: since \minisat\ forms the basis of the majority of modern SAT solvers, an implementation using \minisat\ shows that the format is widely applicable, and provides code which developers can easily incorporate into a large number of existing solvers. \cadical, on the other hand, is a modern solver which employs a wide range of sophisticated optimizations. A successful conversion of \cadical\ shows that the technology is scalable, and is not limited to toy examples.

As mentioned in Section~\ref{sec:frat}, the main solver modifications required for \FRAT\ production are inclusions of clause IDs, finalization steps, and \LRAT\ proof traces. The provision of IDs requires some non-trivial modification as many solvers, including \cadical\ and \minisat, do not natively keep track of clause IDs, and \DRAT\ proofs use literal lists up to permutation for clause identity. In \cadical, we added IDs to all clauses, leading to 8 bytes overhead per clause. Additionally, unit clauses are tracked separately, and ensuring proper ID tracking for unit clauses resulted in some added code complexity. In \minisat, we achieved 0 byte overhead by using the pointer value of clauses as their ID, with unit clauses having computed IDs based on the literal. This requires the use of relocation steps during garbage collection. The output of finalization steps requires identifying the active set from the solver state, which can be subtle depending on the solver architecture, but is otherwise a trivial task assuming knowledge of the solver.

\LRAT\ trace production is the heart of the work, and requires the solver to justify each addition step. This modification is relatively easier to apply to \minisat, as it only adds clauses in a few places, and already tracks the “reasons” for each literal in the current assignment, which makes the proof trace straightforward. In contrast, \cadical\ has over 30 ways to add clauses; in addition to the main CDCL loop, there are various in-processing and optimization passes that can create new clauses.

To accommodate this complexity, we leverage the flexibility of the \FRAT\ format which allows optional hints to focus on the most common clause addition steps, to reap the majority of runtime advantage with only a few changes. The \FRAT\ elaborator falls back on the standard elaboration-by-unit propagation when proofs are not provided, so future work can add more proofs to \cadical\ without any changes to the toolchain.

To maximize the efficacy of the modification, we used a simple method to find places to add proofs. In the first pass, we added support for clause ID tracking and finalization, and changing the output format to \FRAT\ syntax. Since \cadical\ was already producing \DRAT\ proofs, we can easily identify the addition and removal steps and replace them with \texttt{a} and \texttt{d} steps. Once this is done, \cadical\ is producing valid \FRAT\ files which can pass through the elaborator and get \LRAT\ results, but it will be quite slow since the \FRAT\ elaborator is essentially acting as a less-optimized version of \DRATtrim\ at this point.

We then find all code paths that lead to an \texttt{a} step being emitted, and add an extra call to output a step of the form $\mathtt{t}\ \langle todo\_id\rangle\ \mathtt{0}$, where $\langle todo\_id\rangle$ is some unique identifier of this position in the code. These steps are ignored by the \FRAT\ elaborator, but we can use them to identify the hottest code paths by running the solver on benchmarks and counting how many \texttt{t} steps of each kind appear.

The basic idea is that elaborating a step that has a proof is much faster than elaborating a step that doesn't, but the distribution of code paths leading to add steps is highly skewed, so adding proofs to to the top 3 or 4 paths already decreases the elaboration time by over 70\%. At the time of writing, about one third of \cadical\ code paths are covered, and median elaboration time is about 15\% that of \DRATtrim\ (see Section~\ref{sec:results}). (This is despite the fact that our elaborator could stand to improve on low level optimizations, and runs about twice as slow as \DRATtrim\ when no proofs are provided.)

\section{Elaboration}

\begin{algorithm}[tb]
\caption{First pass (elaboration): \FRAT\ to elaborated reversed \FRAT}%
\label{alg:first-pass}
\begin{algorithmic}[1]
\Function{Elaborate}{$cert$}
  \State{$F\gets\emptyset,\ revcert\gets[]$}\Comment{$F$ is a map ID $\to$ clause with a \textbf{bool} marking}
  \For{$step$ \textbf{in} $\mathrm{reverse}(cert)$}
    \Case{$step$}
      \State{$\mathtt{o}(i, C)\Rightarrow$}\Indent
        \State{$C'\gets F.\mathrm{remove}(i)$;\ \textbf{assert} $C'\simeq C$}
        \State{\textbf{if} $C'.\mathrm{marked}$ \textbf{then} $revcert\gets revcert,\mathtt{o}(i,C)$}
      \EndIndent
      \State{$\mathtt{a}(i, C, \omega^?, proof^?)\Rightarrow$}\Indent
        \State{$C'\gets F.\mathrm{remove}(i)$;\ \textbf{assert} $C'\simeq C$}
        \If{$C'.\mathrm{marked}$}
          \State{$steps'\gets{}$\textbf{case} $proof^?$ \textbf{of}}\Indent
            \State{$\varepsilon\Rightarrow\Call{ProvePR}{F,C,\omega^?}$}
            \State{$\mathtt{l}(steps)\Rightarrow\Call{CheckHint}{F,C,\omega^?,steps}$}
          \EndIndent
          \For{$j$ \textbf{in} $\{j\mid \pm j\in steps'\}$}
            \If{$\neg F_j.\mathrm{marked}$}
              \State{$F_j.\mathrm{marked} \gets \mathbf{true}$}
              \State{$revcert\gets revcert,\mathtt{d}(step,F_j)$}
            \EndIf
          \EndFor
          \State{$revcert\gets revcert,\mathtt{a}(i,C,\mathtt{l}(steps'))$}
        \EndIf
      \EndIndent
      \State{$\mathtt{d}(i, C)\Rightarrow F.\mathrm{insert}(i, C, \mathrm{marked}{:}\ \mathbf{false})$}
      \State{$\mathtt{f}(i, C)\Rightarrow F.\mathrm{insert}(i, C, \mathrm{marked}{:}\ C=\bot)$}
      \State{$\mathtt{r}(R)\Rightarrow$}\Indent
        \State{$R'\gets \{(s,t)\in R\mid \exists x. (t, x)\in F\}$}
        \State{$F\gets F-\{(t,F_t)\mid (s,t)\in R'\}+\{(s,F_t)\mid (s,t)\in R'\}$}
        \State{$revcert\gets revcert,\mathtt{r}(R')$}
      \EndIndent
    \EndCase
  \EndFor
  \State{\Return $revcert$}
\EndFunction
\end{algorithmic}
\end{algorithm}

\begin{algorithm}[tb]
\caption{Second pass (renumbering): elaborated reversed \FRAT\ to \LRAT}%
\label{alg:second-pass}
\begin{algorithmic}[1]
\Function{Renumber}{$F_\mathrm{orig},revcert$}
  \State{$M\gets\emptyset,\ k\gets |F_\mathrm{orig}|,\ lrat\gets []$}\Comment{$M$ is a map ID $\to$ ID}
  \For{$step$ \textbf{in} $\mathrm{reverse}(revcert)$}
    \Case{$step$}
      \State{$\mathtt{o}(i, C)\Rightarrow$ find $j$ such that $C\simeq (F_\mathrm{orig})_j$;\ $M.\mathrm{insert}(i,j)$}
      \State{$\mathtt{a}(i, C, \omega^?, \mathtt{l}(steps))\Rightarrow$}\Indent
        \State{$k \gets k+1;\ M.\mathrm{insert}(i,k)$}
        \State{$lrat\gets lrat,\mathtt{add}(k,C,\omega^?,[\pm M_i\mid \pm i\in steps])$}
        \State{\textbf{if} $C=\bot$ \textbf{then} \Return $lrat$}
      \EndIndent
      \State{$\mathtt{d}(i, C)\Rightarrow lrat\gets lrat,\mathtt{del}(k,M.\mathrm{remove}(i))$}
      \State{$\mathtt{r}(R)\Rightarrow M\gets M-\{(s,M_s)\mid (s,t)\in R\}+\{(t,M_s)\mid (s,t)\in R\}$}
    \EndCase
  \EndFor
  \State{\textbf{assert false}}\Comment{no proof of $\bot$ found}
\EndFunction
\end{algorithmic}
\end{algorithm}

The main tasks of the \FRAT-to-\LRAT\ elaborator\footnote{The elaborator used for this paper can be found at \url{https://github.com/digama0/frat/tree/tacas}.} are provision of missing RUP step hints, elimination of irrelevant clause additions, and re-labeling clauses with new IDs. These tasks are performed in two separate `passes' over files, writing and reading directly to disk (so the entire proof is never in memory at once). In the first pass, the elaborator reads the \FRAT\ file and produces a temporary file (which by default is stored on disk, but can also be configured to be stored in memory). The temporary file is essentially the original \FRAT\ file with the steps put in reverse order, while satisfying the following additional conditions:
\begin{itemize}
    \item All \texttt{a} steps have annotations.
    \item Every clause introduced by an \texttt{o}, \texttt{a}, or \texttt{r} step ultimately contributes to the proof of $\bot$. Note that we consider an \texttt{r} step as using an old clause with the old ID and introducing a new clause with the new ID\@.
    \item There are no \texttt{f} steps.
\end{itemize}

\noindent
Algorithm~\ref{alg:first-pass} shows the pseudocode of the first pass, \textsc{Elaborate}($cert$). Here, $cert$ is the \FRAT\ proof obtained from the SAT solver, and the pass works by iterating over its steps in reverse order, producing the temporary file $revcert$. The map $F$ maintains the active formula as a map with unique IDs for each clause (double inserts and removes to $F$ are always error conditions), and the effect of each step is replayed backwards to reconstruct the solver's state at the point each step was produced.

\begin{itemize}
  \item All \texttt{d} or \texttt{f} clauses are immediately inserted to $F$, but (with the exception of the empty clause) are marked as not necessarily required for the proof, and the \texttt{d} step is deferred until just before its first use (or rather, just after the last use).
  \item \textsc{ProvePR}($F,C,\omega^?$), not given here, checks that $C$ has property \textsf{RAT} with respect to $F$, and produces a step list in \LRAT\ format (where positive numbers are clause references in a unit propagation proof, and negative numbers are used in RAT steps, indicating the clauses to resolve against). When a witness $\omega$ is provided (not empty), it checks that $C$ is \textsf{PR} with respect to $F$ and the witness $\omega$. 
  \item \textsc{CheckHint}($F,C,\omega^?,steps$) does the same thing, but it has been given a candidate proof, $steps$. It will check that $steps$ is a valid proof, and if so, returns it, but the steps in the unit propagation proof may be out of order (in which case they are reordered to \LRAT\ conformity), and if the given proof is not salvageable, it falls back on \textsc{ProveRAT}($F,C$) to construct the proof.
\end{itemize}

\noindent
In the second pass, \textsc{Renumber}($F_\mathrm{orig},revcert$) reads the input DIMACS file and the temporary file from the first pass, and produces the final result in \LRAT\ format. Not much checking happens in this pass, but we ensure that the \texttt{o} steps in the \FRAT\ file actually appear (up to permutation) in the input. The state that is maintained in this pass is a list of all active clause IDs, and the corresponding list of \LRAT\ IDs (in which original steps are always numbered sequentially in the file, and add/delete steps use a monotonic counter that is incremented on each addition step).

The resulting \LRAT\ file can then be verified by any of the verified \LRAT\ checkers~\cite{wetzler2013mechanical}, or by the built-in \LRAT\ checker included in our \FRAT\ toolchain, with one caveat: if the input \FRAT\ file includes PR steps described in Section~\ref{sec:pr}, the output will be an \LPR\ file which must be checked using an \LPR\ checker like \textsf{cake\_lpr}~\cite{tan2021cake_lpr}. This usually happens when the \FRAT\ file was generated by converting from a \DPR\ file. To avoid this, run the \FRAT\ toolchain's \texttt{from-pr} subcommand on the original \DPR\ file, which produces a PR-free \FRAT\ output by translating PR steps into a sequence of RAT steps. Then the output can be elaborated into the \LRAT\ format.

The 2-pass algorithm is used in order to optimize memory usage. The result of the first pass is streamed out so that the intermediate elaboration result does not have to be stored in memory simultaneously. Once the temporary file is streamed out, we need at least one more pass to reverse it (even if the labels did not need renumbering) since its steps are in reverse order.

\section{Test results}\label{sec:results}

\subsection{Testing environment}

 All tests in this section were performed on Amazon EC2 \texttt{r5a.xlarge} instances, running Ubuntu Server 20.04 LTS on 2.5 GHz AMD EPYC 7000 processors with 32 GB RAM and 512 GB SSD\@.

\subsection{FRAT vs. DRAT}\label{sec:frat-vs-drat}

We performed benchmarks comparing our \FRAT\ toolchain (modified \cadical\ + \FRAT-to-\LRAT\ elaborator written in Rust) against the \DRAT\ toolchain (standard \cadical\ + \DRATtrim) and measured their execution times, output file sizes, and peak memory usages while solving SAT instances in the DIMACS format and producing their \LRAT\ proofs.

The instances used in the benchmark were chosen by selecting all 97 instances for which default-mode \cadical\ returned `UNSAT' in the 2019 SAT Race results. One of these instances was excluded because \DRATtrim\ exhausted the available 32GB memory and failed during elaboration. Although this instance was not used for comparisons below, we note that it offers further evidence of the \FRAT\ toolchain's efficient use of memory, since the \FRAT-to-\LRAT\ elaboration of this instance succeeded on the same system. The remaining 96 instances were used for performance comparison of the two toolchains.\footnote{CSV files of detailed benchmark results can be found at \url{https://github.com/digama0/frat/blob/lmcs/benchmarks}.}

Figures~\ref{fig:time-chart} and~\ref{fig:mem-chart} show the time and memory measurements from the benchmark. We can see from Figure~\ref{fig:time-chart} that the \FRAT\ toolchain is significantly faster than \DRAT\ toolchain. Although the modified \cadical\ tends to be slightly (6\%) slower than standard \cadical, that overhead is more than compensated by a median 84\% decrease in elaboration time (the sum over all instances are 1700.47 s in the \DRAT\ toolchain vs. 381.70 s in the \FRAT\ toolchain, so the average is down by 77\%). If we include the time of the respective solvers, the execution time of the \FRAT\ + modified \cadical\ toolchain is 53.6\% that of the \DRAT\ + \cadical\ toolchain on median. The difference in the toolchains' time budgets is clear: the \DRAT\ toolchain spends 42\% of its time in solving and 58\% in elaboration, while \FRAT\ spends 85\% on solving and only 15\% on elaboration.

\begin{figure}[tb]
\centering
\begin{tikzpicture}[scale = 1.0]
  \begin{axis}
    [
      mark options={scale=1.0},
      grid=both,
      grid style={black!10},
      legend style={at={(0.97,0.47)}},
      legend cell align={left},
      x post scale=1.5,
      y post scale=1.0,
      xlabel=Time (minutes),
      ylabel=Number of instances,
      xtick={0, 60000 * 100, 60000 * 200, 60000 * 300, 60000 * 400},
      xticklabels={0, 100, 200, 300, 400},
      xmin=0,
      xmax=25000000,
      ymin=0,
      ymax=100,
      scaled x ticks=false
    ]

    \addplot[color=frat, mark=triangle] coordinates { (16333, 1) (17294, 2) (32110, 3) (44166, 4) (45816, 5) (49950, 6) (100685, 7) (132634, 8) (137582, 9) (149331, 10) (191251, 11) (197363, 12) (227674, 13) (232756, 14) (237368, 15) (284560, 16) (314949, 17) (345988, 18) (362950, 19) (370058, 20) (371596, 21) (372863, 22) (373593, 23) (420457, 24) (426583, 25) (444393, 26) (446406, 27) (448092, 28) (449626, 29) (458215, 30) (459801, 31) (465667, 32) (467863, 33) (475753, 34) (477673, 35) (508981, 36) (513462, 37) (596066, 38) (598843, 39) (601976, 40) (664992, 41) (665186, 42) (668169, 43) (685335, 44) (794965, 45) (883130, 46) (1014042, 47) (1032156, 48) (1379650, 49) (1380392, 50) (1381744, 51) (1400712, 52) (1450601, 53) (1515184, 54) (1571369, 55) (1583557, 56) (1650244, 57) (1788852, 58) (1790296, 59) (1809598, 60) (1832943, 61) (1906088, 62) (1952091, 63) (2000764, 64) (2067621, 65) (2071594, 66) (2112794, 67) (2175029, 68) (2220835, 69) (2331255, 70) (2354506, 71) (2371693, 72) (2503835, 73) (2550500, 74) (2620213, 75) (2652847, 76) (2663433, 77) (2692626, 78) (2698860, 79) (2733108, 80) (2746759, 81) (2850217, 82) (2889163, 83) (2940799, 84) (3121336, 85) (3158900, 86) (3224996, 87) (3265341, 88) (3566252, 89) (3759866, 90) (3802622, 91) (3866602, 92) (4353634, 93) (6133480, 94) (8274426, 95) (16309812, 96) };

    \addplot[color=drat, mark=triangle] coordinates { (14730, 1) (19280, 2) (31070, 3) (48060, 4) (52250, 5) (130090, 6) (148040, 7) (176860, 8) (226870, 9) (229950, 10) (274060, 11) (275960, 12) (323990, 13) (330870, 14) (354050, 15) (367420, 16) (367540, 17) (372720, 18) (415590, 19) (436990, 20) (484470, 21) (512830, 22) (542040, 23) (553010, 24) (557780, 25) (672540, 26) (706810, 27) (751780, 28) (826990, 29) (946760, 30) (1012320, 31) (1013730, 32) (1013910, 33) (1027320, 34) (1047350, 35) (1073790, 36) (1075930, 37) (1093920, 38) (1149510, 39) (1157640, 40) (1191660, 41) (1194520, 42) (1311390, 43) (1356230, 44) (1370870, 45) (1446730, 46) (1532960, 47) (1701370, 48) (1762490, 49) (2046200, 50) (2062130, 51) (2239970, 52) (2277620, 53) (2590730, 54) (2711760, 55) (2929410, 56) (2958090, 57) (2990310, 58) (3007890, 59) (3033470, 60) (3073310, 61) (3113520, 62) (3181030, 63) (3328540, 64) (3348860, 65) (3402750, 66) (3429410, 67) (3476310, 68) (3707970, 69) (3707980, 70) (3935330, 71) (4042780, 72) (4125460, 73) (4196550, 74) (4753380, 75) (4950670, 76) (5158800, 77) (5407320, 78) (5420200, 79) (5467380, 80) (5481920, 81) (5545060, 82) (5873980, 83) (5881320, 84) (5971800, 85) (5988760, 86) (6129090, 87) (6215490, 88) (6847710, 89) (7414480, 90) (7528910, 91) (7650190, 92) (9104150, 93) (9200160, 94) (14233070, 95) (24157620, 96) };

    \addplot[color=fratlight, mark=x] coordinates { (1830, 1) (2520, 2) (3140, 3) (3860, 4) (6780, 5) (6970, 6) (15840, 7) (19130, 8) (27950, 9) (32080, 10) (35320, 11) (36290, 12) (37110, 13) (38530, 14) (38670, 15) (39950, 16) (40400, 17) (47390, 18) (49260, 19) (52580, 20) (55590, 21) (58840, 22) (60300, 23) (64780, 24) (65720, 25) (66960, 26) (66970, 27) (71540, 28) (75370, 29) (75920, 30) (76740, 31) (82190, 32) (89710, 33) (95750, 34) (96710, 35) (97180, 36) (97290, 37) (100780, 38) (101200, 39) (102490, 40) (110420, 41) (110840, 42) (111290, 43) (113180, 44) (116220, 45) (116350, 46) (125150, 47) (127800, 48) (129350, 49) (138580, 50) (138700, 51) (140600, 52) (148320, 53) (155830, 54) (156870, 55) (158630, 56) (177920, 57) (181130, 58) (183810, 59) (188150, 60) (189140, 61) (196050, 62) (201600, 63) (207320, 64) (208980, 65) (211000, 66) (213440, 67) (257800, 68) (265760, 69) (278690, 70) (294820, 71) (311610, 72) (314430, 73) (329490, 74) (332720, 75) (340630, 76) (352650, 77) (365920, 78) (413310, 79) (417390, 80) (459870, 81) (463030, 82) (513490, 83) (542570, 84) (553500, 85) (577040, 86) (737190, 87) (749620, 88) (810420, 89) (829180, 90) (846880, 91) (886080, 92) (1811960, 93) (1838200, 94) (2326170, 95) (12028200, 96) };

    \addplot[color=dratlight, mark=x] coordinates { (3220, 1) (3400, 2) (3460, 3) (6360, 4) (13510, 5) (17410, 6) (33430, 7) (39300, 8) (46830, 9) (54460, 10) (74890, 11) (80450, 12) (131440, 13) (136950, 14) (141410, 15) (146560, 16) (148040, 17) (163250, 18) (186640, 19) (189560, 20) (214860, 21) (215330, 22) (234140, 23) (236420, 24) (307200, 25) (323680, 26) (324540, 27) (328800, 28) (348620, 29) (369620, 30) (390180, 31) (419050, 32) (443360, 33) (513600, 34) (587570, 35) (597530, 36) (604850, 37) (623770, 38) (646900, 39) (652800, 40) (659460, 41) (662050, 42) (677420, 43) (683240, 44) (712550, 45) (718790, 46) (722120, 47) (777920, 48) (855940, 49) (873590, 50) (897730, 51) (914730, 52) (931910, 53) (975680, 54) (1124840, 55) (1207440, 56) (1271780, 57) (1360200, 58) (1500190, 59) (1508140, 60) (1559060, 61) (1600120, 62) (1687280, 63) (1796370, 64) (1901760, 65) (2082130, 66) (2100220, 67) (2203590, 68) (2263530, 69) (2296180, 70) (2331340, 71) (2346210, 72) (2379530, 73) (2415270, 74) (2592410, 75) (2734030, 76) (2838750, 77) (2873700, 78) (2892000, 79) (2896680, 80) (3016000, 81) (3101130, 82) (3157860, 83) (3387560, 84) (3460060, 85) (3527540, 86) (3543700, 87) (3662490, 88) (4199770, 89) (4560500, 90) (5187050, 91) (5291770, 92) (5684140, 93) (6456070, 94) (11304700, 95) (18875330, 96) };

    \legend{
      \scriptsize \textsf{FRAT total},
      \scriptsize \textsf{DRAT total},
      \scriptsize \textsf{FRAT elab},
      \scriptsize \textsf{DRAT elab}
    }

  \end{axis}
\end{tikzpicture}
\caption{\FRAT\ vs. \DRAT\ time comparison. Each datapoint of `\FRAT\ total' (resp. `\DRAT\ total') shows the number of instances whose \LRAT\ proofs could be generated by the \FRAT\ (resp. \DRAT) toolchain within the given time limit. Each datapoint of `\FRAT\ elab' (resp. `\DRAT\ elab') shows the number of instances whose  \FRAT\ (resp. \DRAT) proofs could be elaborated into \LRAT\ proofs within the given time limit.}%
\label{fig:time-chart}
\end{figure}
Figure~\ref{fig:mem-chart} shows a dramatic difference in peak memory usage between the \FRAT\ and \DRAT\ toolchains. On median, the \FRAT\ toolchain used only 5.4\% as much peak memory as \DRAT\@. (The average is 318.62 MB, which is 11.98\% that of the \DRAT\ toolchain's 2659.07 MB, but this is dominated by the really large instances. The maximum memory usage was 2.99 GB for \FRAT\ and 21.5 GB for \DRAT, but one instance exhausted the available 32 GB in \DRAT\ and is not included in this figure.) This result is in agreement with our initial expectations: the \FRAT\ toolchain's 2-pass elaboration method allows it to limit the number of clauses held in memory to the size of the active set used by the solver, whereas the \DRAT\ toolchain loads all clauses in a \DRAT\ file into memory at once during elaboration. This difference suggests that the \FRAT\ toolchain can be used to verify instances that would be unverifiable with the \DRAT\ toolchain due to requiring more memory than the system limit.

There were no noticeable differences in the sizes or verification times of \LRAT\ proofs produced by the two toolchains. On average, \LRAT\ proofs produced by the \FRAT\ toolchain were 1.873\% smaller and 3.314\% faster\footnote{One instance was omitted from the LRAT verification time comparison due to what seems to be a bug in the standard LRAT checker included in \DRATtrim. Detailed information regarding this instance can be found at \url{https://github.com/digama0/frat/blob/lmcs/benchmarks/README.md}.} to check than those from the \DRAT\ toolchain.

\begin{figure}[tb]
\centering
\begin{tikzpicture}[scale = 1.0]
  \begin{axis}
    [
      mark options={scale=1.0},
      grid=both,
      grid style={black!10},
      legend style={at={(0.97,0.47)}},
      legend cell align={left},
      x post scale=1.5,
      y post scale=1.0,
      xlabel=Peak memory usage (GB),
      ylabel=Number of instances,
      xtick={0, 5000000, 10000000, 15000000, 20000000},
      xticklabels={0, 5, 10, 15, 20},
      xmin=0,
      xmax=22000000,
      ymin=0,
      ymax=100,
      scaled x ticks=false
    ]

    \addplot[color=frat, mark=triangle] coordinates {
      (18448, 1) (22836, 2) (23344, 3) (23592, 4) (27272, 5) (28260, 6) (30164, 7) (31056, 8) (33844, 9) (34844, 10) (35216, 11) (35272, 12) (35368, 13) (36800, 14) (37276, 15) (37572, 16) (37672, 17) (37728, 18) (38700, 19) (38884, 20) (40468, 21) (40728, 22) (40816, 23) (43000, 24) (43188, 25) (43724, 26) (45900, 27) (47064, 28) (47328, 29) (49432, 30) (53620, 31) (57896, 32) (59052, 33) (61204, 34) (61396, 35) (62904, 36) (65152, 37) (69600, 38) (70404, 39) (70832, 40) (73128, 41) (76944, 42) (78900, 43) (93804, 44) (98184, 45) (99184, 46) (101780, 47) (102772, 48) (103556, 49) (113304, 50) (113640, 51) (119664, 52) (121076, 53) (121128, 54) (137516, 55) (145036, 56) (165736, 57) (169164, 58) (173924, 59) (179996, 60) (180272, 61) (180388, 62) (180488, 63) (193036, 64) (197116, 65) (215816, 66) (216396, 67) (216852, 68) (217516, 69) (262624, 70) (266884, 71) (281304, 72) (284828, 73) (318004, 74) (318436, 75) (413708, 76) (419204, 77) (421412, 78) (517720, 79) (521500, 80) (537356, 81) (551840, 82) (624524, 83) (633632, 84) (687388, 85) (780372, 86) (955280, 87) (959388, 88) (1027928, 89) (1198392, 90) (1302628, 91) (1391524, 92) (1637144, 93) (2615712, 94) (2757880, 95) (2997084, 96)
    };

    \addplot[color=drat, mark=triangle] coordinates {
      (109304, 1) (123008, 2) (127564, 3) (130920, 4) (160732, 5) (182440, 6) (254924, 7) (359512, 8) (432560, 9) (491876, 10) (520228, 11) (550360, 12) (586740, 13) (589060, 14) (602948, 15) (652264, 16) (664860, 17) (677580, 18) (720828, 19) (736404, 20) (761396, 21) (798872, 22) (823740, 23) (858792, 24) (877560, 25) (961016, 26) (1004572, 27) (1083380, 28) (1105724, 29) (1147512, 30) (1190784, 31) (1194684, 32) (1196524, 33) (1201420, 34) (1219540, 35) (1249840, 36) (1270992, 37) (1298568, 38) (1311192, 39) (1374472, 40) (1396720, 41) (1401228, 42) (1416420, 43) (1529660, 44) (1530404, 45) (1587860, 46) (1588004, 47) (1609060, 48) (1626968, 49) (1631572, 50) (1700308, 51) (1741836, 52) (1799448, 53) (1813384, 54) (1826132, 55) (1962852, 56) (2119596, 57) (2168532, 58) (2241424, 59) (2330216, 60) (2350824, 61) (2417808, 62) (2425996, 63) (2461316, 64) (2462660, 65) (2671284, 66) (2674908, 67) (2703952, 68) (2709932, 69) (2863072, 70) (2999500, 71) (3007944, 72) (3011976, 73) (3305972, 74) (3348684, 75) (3469628, 76) (3487588, 77) (3567712, 78) (3624304, 79) (3787004, 80) (3854756, 81) (4452216, 82) (4518488, 83) (4529796, 84) (4857920, 85) (5391492, 86) (5952412, 87) (6057448, 88) (6429576, 89) (6561056, 90) (8666116, 91) (9499304, 92) (9815944, 93) (11371892, 94) (16788356, 95) (21545744, 96)
    };

    \legend{
      \scriptsize \textsf{FRAT},
      \scriptsize \textsf{DRAT}
    }

  \end{axis}
\end{tikzpicture}
  \caption{\FRAT\ vs. \DRAT\ peak memory usage comparison. Each datapoint of `\FRAT' (resp. `\DRAT') shows the number of instances whose \LRAT\ proofs could be generated by the \FRAT\ (resp. \DRAT) toolchain within the given peak memory usage limit.}%
  \label{fig:mem-chart}
\end{figure}
One minor downside of the \FRAT\ toolchain is that it requires the storage of a temporary file during elaboration, but we do not expect this to be a problem in practice since the temporary file is typically much smaller than either the \FRAT\ or \LRAT\ file. In our test cases, the average temporary file size was 28.68\% and 47.60\% that of \FRAT\ and \LRAT\ files, respectively. In addition, users can run the elaborator with the \texttt{-m} option to bypass temporary files and write the temporary data to memory instead, which further improves performance but foregoes the memory conservation that comes with 2-pass elaboration.

The modified \cadical\ used for the benchmark is a prototype, and some of its weaknesses show in the data. The general pattern we observed is that on problems for which the predicted \cadical\ code paths were taken, the generated files have a large number of hints and the elaboration time is negligible (the ``\FRAT\ elab'' line in Fig.~\ref{fig:time-chart}); but on problems which make use of the more unusual in-processing operations, many steps with no hints are given to the elaborator, and performance becomes comparable to \DRATtrim. For solver developers, this means that there is a very direct relationship between proof annotation effort and mean solution + elaboration time.

\subsection{\FRATZ\ vs. \DRAT}\label{sec:fratz-vs-drat}

The performance improvement of the \FRAT\ toolchain relative to the \DRAT\ toolchain comes from two main sources, (1) the optional hints and (2) the 2-pass elaboration model. Since the two optimizations are independent of each other, it would be interesting to investigate how each of them affect elaboration performance in isolation. In particular, we'd want to know how the \FRAT\ toolchain performs when using the 2-pass elaboration on completely unannotated \FRAT\ proofs (which, for convenience, we refer to as `\FRATZ'), since this is a realistic scenario for developers who find it difficult to modify a solver for hint generation and seek to use the \FRAT\ format for the memory efficiency advantage, or to interface with other \FRAT-based tools.

In order to perform this new test, we had to first update the elaborator with new optimizations. The original version of the elaborator used for the tests in Section~\ref{sec:frat-vs-drat} was designed with the assumption that most steps will be annotated, falling back to naive searches and becoming prohibitively slow whenever a significant proportion of steps came without hints. This problem was fixed by porting relevant
optimization techniques from \DRATtrim\ to the elaborator, putting their elaboration speeds (roughly) on par. Using this new version of the elaborator, we performed another benchmark that compares the elaboration time and memory usage of \FRATZ\ and \DRAT\ files. The files used for the benchmark were identical to that of Section~\ref{sec:frat-vs-drat}, except that all hints were removed from the FRAT files using the \texttt{strip-frat} subcommand of the elaborator. The results are shown in Figures~\ref{fig:time-chart-2} and~\ref{fig:mem-chart-2}.

\begin{figure}[tb]
\centering
\begin{tikzpicture}[scale = 1.0]
  \begin{axis}
    [
      mark options={scale=1.0},
      grid=both,
      grid style={black!10},
      legend style={at={(0.97,0.47)}},
      legend cell align={left},
      x post scale=1.5,
      y post scale=1.0,
      xlabel=Time (minutes),
      ylabel=Number of instances,
      xtick={0, 60000 * 100, 60000 * 200, 60000 * 300, 60000 * 400, 60000 * 500},
      xticklabels={0, 100, 200, 300, 400, 500},
      xmin=0,
      xmax=30000000,
      ymin=0,
      ymax=100,
      scaled x ticks=false
    ]

    \addplot[color=frat, mark=x] coordinates { (2300, 1) (2560, 2) (2700, 3) (8150, 4) (13810, 5) (25780, 6) (31580, 7) (54670, 8) (58180, 9) (66750, 10) (97310, 11) (113210, 12) (149420, 13) (164490, 14) (174660, 15) (189610, 16) (194250, 17) (258520, 18) (262500, 19) (272640, 20) (279000, 21) (296880, 22) (341490, 23) (362690, 24) (366670, 25) (385200, 26) (405720, 27) (407450, 28) (419580, 29) (445610, 30) (464320, 31) (489770, 32) (627020, 33) (670310, 34) (690370, 35) (693050, 36) (705900, 37) (721180, 38) (765520, 39) (767200, 40) (769080, 41) (800300, 42) (840470, 43) (871620, 44) (881460, 45) (884550, 46) (891650, 47) (968190, 48) (1056120, 49) (1082190, 50) (1184350, 51) (1191830, 52) (1256890, 53) (1306020, 54) (1333040, 55) (1408010, 56) (1463390, 57) (1541850, 58) (1714750, 59) (1818910, 60) (1856910, 61) (1939510, 62) (1964000, 63) (2015260, 64) (2072900, 65) (2199540, 66) (2249300, 67) (2388110, 68) (2433670, 69) (2527420, 70) (2543920, 71) (2556290, 72) (2771050, 73) (2907100, 74) (2982180, 75) (3149550, 76) (3173380, 77) (3207910, 78) (3347520, 79) (3566090, 80) (3746960, 81) (4001210, 82) (4017280, 83) (4516360, 84) (4664140, 85) (5166720, 86) (5774250, 87) (5888000, 88) (5992200, 89) (6063280, 90) (8030890, 91) (8109340, 92) (10492400, 93) (10649490, 94) (20790170, 95) (28476490, 96) };

    \addplot[color=drat, mark=x] coordinates { (3220, 1) (3400, 2) (3460, 3) (6360, 4) (13510, 5) (17410, 6) (33430, 7) (39300, 8) (46830, 9) (54460, 10) (74890, 11) (80450, 12) (131440, 13) (136950, 14) (141410, 15) (146560, 16) (148040, 17) (163250, 18) (186640, 19) (189560, 20) (214860, 21) (215330, 22) (234140, 23) (236420, 24) (307200, 25) (323680, 26) (324540, 27) (328800, 28) (348620, 29) (369620, 30) (390180, 31) (419050, 32) (443360, 33) (513600, 34) (587570, 35) (597530, 36) (604850, 37) (623770, 38) (646900, 39) (652800, 40) (659460, 41) (662050, 42) (677420, 43) (683240, 44) (712550, 45) (718790, 46) (722120, 47) (777920, 48) (855940, 49) (873590, 50) (897730, 51) (914730, 52) (931910, 53) (975680, 54) (1124840, 55) (1207440, 56) (1271780, 57) (1360200, 58) (1500190, 59) (1508140, 60) (1559060, 61) (1600120, 62) (1687280, 63) (1796370, 64) (1901760, 65) (2082130, 66) (2100220, 67) (2203590, 68) (2263530, 69) (2296180, 70) (2331340, 71) (2346210, 72) (2379530, 73) (2415270, 74) (2592410, 75) (2734030, 76) (2838750, 77) (2873700, 78) (2892000, 79) (2896680, 80) (3016000, 81) (3101130, 82) (3157860, 83) (3387560, 84) (3460060, 85) (3527540, 86) (3543700, 87) (3662490, 88) (4199770, 89) (4560500, 90) (5187050, 91) (5291770, 92) (5684140, 93) (6456070, 94) (11304700, 95) (18875330, 96) };

    \legend{
      \scriptsize \textsf{FRAT0},
      \scriptsize \textsf{DRAT}
    }

  \end{axis}
\end{tikzpicture}
\caption{
\FRATZ\ vs. \DRAT\ time comparison.
Each datapoint of `\FRATZ' (resp. `\DRAT') shows the number of instances whose \FRATZ\ (resp. \DRAT) proofs could be elaborated into \LRAT\ proofs within the given time limit.
}%
\label{fig:time-chart-2}
\end{figure}

\begin{figure}[tb]
\centering
\begin{tikzpicture}[scale = 1.0]
  \begin{axis}
    [
      mark options={scale=1.0},
      grid=both,
      grid style={black!10},
      legend style={at={(0.97,0.47)}},
      legend cell align={left},
      x post scale=1.5,
      y post scale=1.0,
      xlabel=Peak memory usage (GB),
      ylabel=Number of instances,
      xtick={0, 5000000, 10000000, 15000000, 20000000},
      xticklabels={0, 5, 10, 15, 20},
      xmin=0,
      xmax=22000000,
      ymin=0,
      ymax=100,
      scaled x ticks=false
    ]

    \addplot[color=frat, mark=triangle] coordinates {
      (19100, 1) (24196, 2) (24784, 3) (24868, 4) (29212, 5) (29404, 6) (32316, 7) (34748, 8) (36884, 9) (37288, 10) (37696, 11) (38000, 12) (39732, 13) (40048, 14) (40212, 15) (40604, 16) (41636, 17) (41932, 18) (42208, 19) (44208, 20) (44316, 21) (44452, 22) (47224, 23) (47712, 24) (49084, 25) (49752, 26) (49980, 27) (50364, 28) (54704, 29) (59016, 30) (61856, 31) (63384, 32) (65748, 33) (66860, 34) (68920, 35) (69696, 36) (74628, 37) (78708, 38) (81100, 39) (81336, 40) (84336, 41) (86232, 42) (98976, 43) (106900, 44) (107416, 45) (108024, 46) (112916, 47) (123608, 48) (127420, 49) (129580, 50) (131260, 51) (133244, 52) (137876, 53) (138428, 54) (139608, 55) (149932, 56) (171944, 57) (172100, 58) (180896, 59) (185328, 60) (194588, 61) (195172, 62) (195616, 63) (199628, 64) (207428, 65) (216796, 66) (220904, 67) (224396, 68) (227180, 69) (265268, 70) (269332, 71) (288092, 72) (295104, 73) (318208, 74) (320340, 75) (421308, 76) (427180, 77) (430556, 78) (515804, 79) (542288, 80) (543388, 81) (561464, 82) (631584, 83) (648096, 84) (712324, 85) (790960, 86) (963832, 87) (968676, 88) (1030376, 89) (1242176, 90) (1304996, 91) (1397660, 92) (1625536, 93) (2622760, 94) (2733892, 95) (3180272, 96)

    };

    \addplot[color=drat, mark=triangle] coordinates {
      (109304, 1) (123008, 2) (127564, 3) (130920, 4) (160732, 5) (182440, 6) (254924, 7) (359512, 8) (432560, 9) (491876, 10) (520228, 11) (550360, 12) (586740, 13) (589060, 14) (602948, 15) (652264, 16) (664860, 17) (677580, 18) (720828, 19) (736404, 20) (761396, 21) (798872, 22) (823740, 23) (858792, 24) (877560, 25) (961016, 26) (1004572, 27) (1083380, 28) (1105724, 29) (1147512, 30) (1190784, 31) (1194684, 32) (1196524, 33) (1201420, 34) (1219540, 35) (1249840, 36) (1270992, 37) (1298568, 38) (1311192, 39) (1374472, 40) (1396720, 41) (1401228, 42) (1416420, 43) (1529660, 44) (1530404, 45) (1587860, 46) (1588004, 47) (1609060, 48) (1626968, 49) (1631572, 50) (1700308, 51) (1741836, 52) (1799448, 53) (1813384, 54) (1826132, 55) (1962852, 56) (2119596, 57) (2168532, 58) (2241424, 59) (2330216, 60) (2350824, 61) (2417808, 62) (2425996, 63) (2461316, 64) (2462660, 65) (2671284, 66) (2674908, 67) (2703952, 68) (2709932, 69) (2863072, 70) (2999500, 71) (3007944, 72) (3011976, 73) (3305972, 74) (3348684, 75) (3469628, 76) (3487588, 77) (3567712, 78) (3624304, 79) (3787004, 80) (3854756, 81) (4452216, 82) (4518488, 83) (4529796, 84) (4857920, 85) (5391492, 86) (5952412, 87) (6057448, 88) (6429576, 89) (6561056, 90) (8666116, 91) (9499304, 92) (9815944, 93) (11371892, 94) (16788356, 95) (21545744, 96)
    };

    \legend{
      \scriptsize \textsf{FRAT0},
      \scriptsize \textsf{DRAT}
    }

  \end{axis}
\end{tikzpicture}
  \caption{\FRATZ\ vs. \DRAT\ peak memory usage comparison. Each datapoint of `\FRATZ' (resp. `\DRAT') shows the number of instances whose \FRATZ\ (resp. \DRAT) proofs could be elaborated into \LRAT\ proofs within the given peak memory usage limit. }%
  \label{fig:mem-chart-2}
\end{figure}

Although the elaboration speed advantage is gone without proof hints, \FRATZ\ still trails \DRAT\ closely thanks to the new optimizations. In return for the slightly increased elaboration time, we get steep cuts in memory usage, which has remained virtually unchanged from the \FRAT\ test results. The data shows that the \FRAT\ toolchain can confer significant advantages whenever memory usage is the bottleneck, even if you cannot implement any major modifications for the SAT solver used.

\subsection{\FRAT\ elaborator vs. \DPRtrim}\label{sec:frat-elab-vs-dpr-trim}

In addition to offering performance improvements over the \DRAT\ format and toolchain, the flexibility of the \FRAT\ format allows it to also serve as an alternative to the more recent and expressive \DPR\ format. Since the \DPR\ format is much more nascent than \DRAT, it was difficult to do a direct comparison with two solvers producing equivalent proofs side-by-side. Instead, we performed a proxy comparison by taking the example \DPR\ proofs from the \texttt{pr-proofs} repository\footnote{\url{https://github.com/marijnheule/pr-proofs/}} and producing their \FRAT\ equivalents with the \DPR-to-\FRAT\ converter included in the \FRAT\ elaborator. The elaboration test results using these instances are shown in Figures~\ref{fig:time-chart-3} and~\ref{fig:mem-chart-3}.

\begin{figure}[tb]
	\centering
	\begin{tikzpicture}[scale = 1.0]
	  \begin{axis}
	    [
	      xmode=log,
	      mark options={scale=1.0},
	      grid=both,
	      grid style={black!10},
	      legend style={at={(0.97,0.47)}},
	      legend cell align={left},
	      x post scale=1.5,
	      y post scale=1.0,
	      xlabel=Time (seconds),
	      ylabel=Number of instances,
	      xtick={10, 100, 1000, 10000, 100000, 1000000},
	      xticklabels={0.01, 0.1, 1, 10, 100, 1000},
	      xmin=5,
	      xmax=2000000,
	      ymin=0,
	      ymax=54,
	      scaled x ticks=false
	    ]

	    \addplot[color=frat, mark=x] coordinates {
(10, 25) (10, 26) (10, 27) (10, 28) (10, 29) (10, 30) (20, 31) (20, 32) (20, 33) (30, 34) (40, 35) (40, 36) (80, 37) (100, 38) (140, 39) (420, 40) (770, 41) (1390, 42) (1720, 43) (4650, 44) (6580, 45) (9250, 46) (9400, 47) (26330, 48) (34140, 49) (46150, 50) (52000, 51) (104430, 52) (499840, 53)
	    };

	    \addplot[color=drat, mark=x] coordinates {
(30, 1) (30, 2) (30, 3) (30, 4) (30, 5) (30, 6) (30, 7) (40, 8) (40, 9) (40, 10) (40, 11) (40, 12) (40, 13) (40, 14) (50, 15) (50, 16) (50, 17) (50, 18) (50, 19) (50, 20) (50, 21) (60, 22) (60, 23) (60, 24) (60, 25) (60, 26) (70, 27) (70, 28) (70, 29) (80, 30) (80, 31) (80, 32) (90, 33) (90, 34) (120, 35) (120, 36) (180, 37) (180, 38) (300, 39) (1340, 40) (1390, 41) (2080, 42) (2840, 43) (13300, 44) (14300, 45) (17720, 46) (25300, 47) (67230, 48) (68370, 49) (125100, 50) (224940, 51) (1092610, 52) (1275110, 53)
	    };

	    \legend{
	      \scriptsize \textsf{\FRAT\ elab},
	      \scriptsize \textsf{\DPR-trim}
	    }

	  \end{axis}
	\end{tikzpicture}
	\caption{\FRAT\ elaborator vs. \DPRtrim\ time comparison. Each datapoint of `\FRAT\ elab' (resp. `\DPRtrim') shows the number of instances whose \FRAT\ (resp. \DPR) proofs could be elaborated into \LPR\ proofs within the given time limit. Note that the $x$-axis scale is logarithmic, and the lowest datapoints of the \FRAT\ graph are not shown because their execution times were measured as 0 seconds with the given measurement resolution (0.01 seconds).
	}%
	\label{fig:time-chart-3}
	\end{figure}

\begin{figure}[tb]
\centering
\begin{tikzpicture}[scale = 1.0]
  \begin{axis}
    [
      xmode=log,
      mark options={scale=1.0},
      grid=both,
      grid style={black!10},
      legend style={at={(0.97,0.47)}},
      legend cell align={left},
      x post scale=1.5,
      y post scale=1.0,
      xlabel=Peak memory usage (MB),
      ylabel=Number of instances,
      xtick={3000, 30000, 300000},
      xticklabels={3,  30, 300},
      xmin=2200,
      xmax=320000,
      ymin=0,
      ymax=54,
      scaled x ticks=false
    ]

    \addplot[color=frat, mark=triangle] coordinates {

    (2720, 1) (2740, 2) (2856, 3) (2948, 4) (2964, 5) (2984, 6) (3004, 7) (3008, 8) (3012, 9) (3028, 10) (3040, 11) (3040, 12) (3040, 13) (3072, 14) (3104, 15) (3148, 16) (3148, 17) (3156, 18) (3160, 19) (3168, 20) (3180, 21) (3192, 22) (3196, 23) (3212, 24) (3232, 25) (3236, 26) (3276, 27) (3304, 28) (3312, 29) (3312, 30) (3344, 31) (3360, 32) (3364, 33) (3412, 34) (3424, 35) (3484, 36) (3532, 37) (4036, 38) (4200, 39) (4680, 40) (4924, 41) (6696, 42) (8808, 43) (9724, 44) (9784, 45) (12476, 46) (16344, 47) (20904, 48) (22100, 49) (22760, 50) (27820, 51) (53372, 52) (65644, 53)

    };

    \addplot[color=drat, mark=triangle] coordinates {
(64188, 1) (64200, 2) (64208, 3) (64224, 4) (64228, 5) (64244, 6) (64248, 7) (64248, 8) (64248, 9) (64252, 10) (64260, 11) (64268, 12) (64268, 13) (64272, 14) (64272, 15) (64280, 16) (64280, 17) (64280, 18) (64280, 19) (64284, 20) (64284, 21) (64288, 22) (64292, 23) (64292, 24) (64308, 25) (64312, 26) (64316, 27) (64324, 28) (64324, 29) (64332, 30) (64340, 31) (64348, 32) (64376, 33) (64376, 34) (64376, 35) (64388, 36) (64404, 37) (64588, 38) (64628, 39) (64984, 40) (65180, 41) (65344, 42) (65456, 43) (66320, 44) (66724, 45) (67064, 46) (69160, 47) (71728, 48) (73852, 49) (106628, 50) (109480, 51) (117216, 52) (273740, 53)
    };

    \legend{
      \scriptsize \textsf{\FRAT\ elab},
      \scriptsize \textsf{\DPR-trim}
    }

  \end{axis}
\end{tikzpicture}
  \caption{\FRAT\ elaborator vs. \DPRtrim\ peak memory usage comparison. Each datapoint of `\FRAT\ elab' (resp. `\DPRtrim') shows the number of instances whose \FRAT\ (resp. \DPR) proofs could be elaborated into \LPR\ proofs within the given peak memory usage limit.
 }%
  \label{fig:mem-chart-3}
\end{figure}

As in the \FRAT\ vs. \DRAT\ benchmarks, the elaboration of \FRAT\ proofs require significantly less memory than that of \DPR, which is expected given its 2-pass elaboration model. What's more surprising is the \FRAT\ elaborator's significant speedup over \DPRtrim: unlike in the \FRAT\ vs. \DRAT\ case, the \FRAT\ proofs in this benchmark were generated from the \DPR\ proofs and do not contain any additional information, so the faster speed can be entirely attributed to the elaborator itself. At any rate, the results show that \FRAT\ can be a highly viable alternative to \DPR, at least when relative to the current version of \DPRtrim. In fact, replacing a \DPR\ toolchain with a \FRAT\ one does not even require any solver modification: the \FRAT\ elaborator's \DPR-to-\FRAT\ conversion takes negligible execution time and uses similar amounts of memory compared to the \FRAT-to-\LPR\ elaboration stage, which means we can take any \DPR-producing solver and plug it to the converter to enjoy almost the same performance improvements as using a natively \FRAT-producing solver.

\subsection{\FRAT\ elaborator vs. \PRTODRAT\ + \DPRtrim}\label{sec:frat-elab-vs-pr2drat}

In some cases, it may be desirable to produce an \LRAT\ proof output instead of \LPR, as the former is supported by a wider range of existing tools. The standard \DPR\ toolchain for doing this is \PRTODRAT\ + \DPRtrim, where the former converts \DPR\ input into \DRAT, and the latter elaborates \DRAT\ into \LRAT\ (note that \DPR\ is a superset of \DRAT, so \DPRtrim\ can elaborate \DRAT\ proofs just like \DRATtrim). Alternatively, we can use \FRAT\ as the intermediate format between \DPR\ and \LRAT, using the converter included in the \FRAT\ elaborator for \DPR-to-\FRAT\ conversion. We benchmarked the performances of the two elaboration methods using the 19 example instances from the \PRTODRAT\ repository\footnote{\url{https://github.com/marijnheule/pr2drat}},
whose results are shown in Figures~\ref{fig:time-chart-4} and~\ref{fig:mem-chart-4}.

\begin{figure}[tb]
\centering
\begin{tikzpicture}[scale = 1.0]
  \begin{axis}
    [
      mark options={scale=1.0},
      grid=both,
      grid style={black!10},
      xmode=log,
      legend style={at={(0.97,0.47)}},
      legend cell align={left},
      x post scale=1.5,
      y post scale=1.0,
      xlabel=Time (seconds),
      ylabel=Number of instances,
      xtick={100, 1000, 10000, 100000, 1000000, 10000000},
      xticklabels={0.1, 1, 10, 100, 1000, 10000},
      xmin=20,
      xmax=25000000,
      ymin=0,
      ymax=20,
      scaled x ticks=false
    ]

    \addplot[color=frat, mark=triangle] coordinates {
    (0, 1) (0, 2) (90, 3) (250, 4) (270, 5) (370, 6) (490, 7) (2470, 8) (2730, 9) (10320, 10) (45760, 11) (86770, 12) (183990, 13) (327890, 14) (1105880, 15) (1396020, 16) (1505550, 17) (7133920, 18) (9462130, 19)
    };

    \addplot[color=drat, mark=triangle] coordinates {
    (40, 1) (40, 2) (200, 3) (480, 4) (500, 5) (680, 6) (870, 7) (2900, 8) (4720, 9) (15890, 10) (55930, 11) (187010, 12) (309440, 13) (443330, 14) (2190700, 15) (2403710, 16) (2411770, 17) (12312540, 18) (17617260, 19)
    };

    \legend{
      \scriptsize \textsf{FRAT elab},
      \scriptsize \textsf{PR2DRAT + DPR-trim},
    }

  \end{axis}
\end{tikzpicture}

	\caption{\FRAT\ elaborator vs. \PRTODRAT\ + \DPRtrim\ time comparison. Each datapoint of `\FRAT\ elab' (resp. `\PRTODRAT\ + \DPRtrim') shows the number of instances whose \LRAT\ proofs could be generated by the \FRAT\ elaborator (resp. \PRTODRAT\ and \DPRtrim) within the given time limit. Note that the $x$-axis scale is logarithmic, and the lowest datapoints of the \FRAT\ graph are not shown because their execution times were measured as 0 seconds with the given measurement resolution (0.01 seconds).
	}\label{fig:time-chart-4}

\end{figure}

\begin{figure}[tb]
\centering
\begin{tikzpicture}[scale = 1.0]
  \begin{axis}
    [
      xmode=log,
      mark options={scale=1.0},
      grid=both,
      grid style={black!10},
      legend style={at={(0.97,0.47)}},
      legend cell align={left},
      x post scale=1.5,
      y post scale=1.0,
      xlabel=Peak memory usage (MB),
      ylabel=Number of instances,
      xtick={10000, 100000, 1000000, 10000000},
      xticklabels={10, 100, 1000, 10000},
      xmin=2200,
      xmax=10000000,
      ymin=0,
      ymax=20,
      scaled x ticks=false
    ]

    \addplot[color=frat, mark=triangle] coordinates {
(2868, 1) (2920, 2) (3188, 3) (3396, 4) (3496, 5) (3528, 6) (3584, 7) (4388, 8) (4460, 9) (6668, 10) (8204, 11) (12012, 12) (15144, 13) (18012, 14) (25648, 15) (26436, 16) (42440, 17) (77108, 18) (81872, 19)
    };

    \addplot[color=drat, mark=triangle] coordinates {
(64304, 1) (64312, 2) (65872, 3) (69344, 4) (69688, 5) (71416, 6) (71568, 7) (73460, 8) (73828, 9) (98224, 10) (108812, 11) (138652, 12) (220192, 13) (501324, 14) (502128, 15) (505328, 16) (1420212, 17) (2022832, 18) (6131036, 19)
    };

    \legend{
      \scriptsize \textsf{\FRAT\ elab},
      \scriptsize \textsf{PR2DRAT + DPR-trim}
    }

  \end{axis}
\end{tikzpicture}
  \caption{\FRAT\ elaborator vs. \PRTODRAT\ + \DPRtrim\ peak memory usage comparison. Each datapoint of `\FRAT\ elab' (resp. `\PRTODRAT\ + \DPRtrim') shows the number of instances whose \DPR\ proofs could be elaborated into \LRAT\ proofs by the \FRAT\ elaborator (resp. \PRTODRAT\  and \DPRtrim) within the given peak memory usage limit.
 }%
  \label{fig:mem-chart-4}
\end{figure}

Again, the results follow a similar pattern observed in other benchmarks. Memory usage shows a dramatic improvement by a couple orders of magnitude, while elaboration speedup is less pronounced but still consistent across the board. This shows that the \FRAT\ format and toolchain is useful not only as a \DPR\ replacement, but also for complimenting \DPR\@.

\section{Related works}

As already mentioned, the \FRAT\ format is most closely related to the \DRAT\ format~\cite{heule2016drat}, which it seeks to replace as an intermediate output format for SAT solvers. It is also dependent on the \LRAT\ format and related tools~\cite{lrat}, as the \FRAT\ toolchain targets \LRAT\ as the final output format.

The \textsf{GRAT} format~\cite{lammich2017grat} and toolchain also aims to improve elaboration of SAT unsatisfiability proofs, but takes a different approach from that of \FRAT\@. It retains \DRAT\ as the intermediate format, but uses parallel processing and targets a new final format with more information than \LRAT\ in order to improve overall performance. \textsf{GRAT} also comes with its own verified checker~\cite{lammich2019efficient}.

Specifying and verifying the program correctness of SAT solvers (sometimes called the \textit{autarkic} method, as opposed to the proof-producing \textit{skeptical} method) is a radically different approach to ensuring the correctness of SAT solvers. There have been various efforts to verify nontrivial SAT solvers~\cite{maric2010formal,Shankar,oe2012versat,DBLP:conf/nfm/Fleury19,DBLP:conf/cpp/FleuryBL18}. Although these solvers have become significantly faster, they cannot compete with the (unverified) state-of-the-art solvers. It is also difficult to maintain and modify certified solvers. Proving the correctness of nontrivial SAT solvers can provide new insights about key invariants underlying the used techniques~\cite{DBLP:conf/cpp/FleuryBL18}.

Generally speaking, devising proof formats for automated reasoning tools and augmenting the tools with proof output capability is an active research area. Notable examples outside SAT solving include the LFSC format for SMT solving~\cite{stump2013smt} and the TSTP format for classical first-order ATPs~\cite{sutcliffe2004tstp}. In particular, the recent work on the \textsc{veriT} SMT solver~\cite{barbosa2019scalable} is motivated by similar rationales as that for the \FRAT\ toolchain; the key insight is that a proof production pipeline is often easier to optimize on the solver side than on the elaborator side, as the former has direct access to many types of useful information.

\section{Conclusion}

The test results show that the \FRAT\ format and toolchain made significant performance gains relative to their \DRAT\ equivalents in both elaboration time and memory usage. We take this as confirmation of our initial conjectures that (1) there is a large amount of useful and easily extracted information in SAT solvers that is left untapped by \DRAT\ proofs, and (2) the use of streaming verification is the key to verifying very large proofs that cannot be held in memory at once.

The practical ramification is that, provided that solvers produce well-annotated \FRAT\ proofs, the elaborator is no longer a bottleneck in the pipeline. Typically, when \DRATtrim\ hangs it does so either by taking excessive time, or by attempting to read in an entire proof file at once and exhausting memory (the so-called ``uncheckable'' proofs that can be produced but not verified). But \FRAT-to-\LRAT\ elaboration is typically faster than \FRAT\ production, and the memory consumption of the \FRAT-to-\LRAT\ elaborator at any given point is proportional to the memory used by the solver at the same point in the proof. Since \LRAT\ verification is already efficient, the only remaining limiting factor is essentially the time and memory usage of the solver itself.

In addition to performance, the other main consideration in the design of the \FRAT\ format and toolchain was flexibility of use and extension. The encoding of \FRAT\ files allows them to be read and parsed both backward and forward, and the format can be modified to include more advanced inferences, as we have seen in the example of PR steps. The optional \texttt{l} steps allow SAT solvers to decide precisely when they will provide explicit proofs, thereby promoting a workable compromise between implementation complexity and runtime efficiency. SAT solver developers can begin using the format by producing the most bare-bones \FRATZ\ proofs with no annotations (essentially \DRAT\ proofs with metadata for original/final clauses) and gradually work toward providing more complete hints. We hope that this combination of efficiency and flexibility will motivate performance-minded SAT solver developers to adopt the format and support more robust proof production, which is presently only an afterthought in most SAT solvers.

\bibliographystyle{alphaurl}
\bibliography{references}

\vfill

{\small\medskip\noindent{\bf Open Access} This chapter is licensed under the terms of the Creative Commons\break Attribution 4.0 International License (\url{http://creativecommons.org/licenses/by/4.0/}), which permits use, sharing, adaptation, distribution and reproduction in any medium or format, as long as you give appropriate credit to the original author(s) and the source, provide a link to the Creative Commons license and indicate if changes were made.} 

{\small \spaceskip .28em plus .1em minus .1em The images or other third party material in this chapter are included in the chapter's Creative Commons license, unless indicated otherwise in a credit line to the material.~If material is not included in the chapter's Creative Commons license and your intended\break use is not permitted by statutory regulation or exceeds the permitted use, you will need to obtain permission directly from the copyright holder.} 

\medskip\noindent\includegraphics{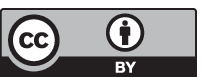}

\end{document}